\documentclass[floatfix,aps,prl,twocolumn,groupedaddress,showpacs,superscriptaddress,preprintnumbers]{revtex4-2}

\usepackage[utf8]{inputenc}
\usepackage{graphicx}
\usepackage{slashed}
\usepackage{color}
\usepackage{amsmath}
\usepackage{amssymb}

\def\wpol{\varepsilon_W}
\def\lrbrace#1{\lbrace#1\rbrace}
\def\nn{\nonumber \\ }
\def\eps{\epsilon}
\newcommand{\udtoWbb}{$u\bar{d}\to W^+b\bar{b}$~}
\def\cI{\mathcal{I}}
\def\cO{\mathcal{O}}

\usepackage{multirow}
\usepackage{tabularx}

\begin{document}

\newcolumntype{C}[1]{>{\hsize=#1\hsize\centering\arraybackslash}X}

\preprint{CAVENDISH-HEP-21/01}

\title{Two-loop QCD corrections to $Wb\bar{b}$ production at hadron colliders}

\author{Simon Badger}
\email{simondavid.badger@unito.it}
\affiliation{
Dipartimento di Fisica and Arnold-Regge Center, Università di Torino, and INFN, Sezione di
Torino, Via P. Giuria 1, I-10125 Torino, Italy
}

\author{Heribertus Bayu Hartanto}
\email{hbhartanto@hep.phy.cam.ac.uk}
\affiliation{
Cavendish Laboratory, University of Cambridge, Cambridge CB3 0HE, United Kingdom
}

\author{Simone Zoia}
\email{simone.zoia@unito.it}
\affiliation{
Dipartimento di Fisica and Arnold-Regge Center, Università di Torino, and INFN, Sezione di
Torino, Via P. Giuria 1, I-10125 Torino, Italy
}

\date{\today}

\begin{abstract}
  We present an analytic computation of the two-loop QCD corrections to
  \udtoWbb for an on-shell $W$-boson using the leading colour and massless bottom quark
  approximations.  We perform an integration-by-parts reduction of the
  unpolarised squared matrix element using finite field reconstruction
  techniques and identify an independent basis of special functions that allows
  an analytic subtraction of the infrared and ultraviolet poles. This basis is
  valid for all planar topologies for five-particle scattering with an
  off-shell leg.
\end{abstract}

\maketitle

\section{Introduction \label{sec:intro}}

The production of a $W$-boson in association with a pair of $b$-quarks at hadron
colliders is of fundamental importance as a background to Higgs
production in association with a vector boson. The process is one of a prioritised list of $2\to 3$ scattering
problems for which higher order corrections are necessary to keep theory in
line with data. These amplitudes are related to a large class of processes
contributing to $pp\to W+2j$ production and the work presented here represents a
significant step towards achieving a complete classification of the missing
two-loop amplitudes.

The process has been studied extensively at next-to-leading order~(NLO)~\cite{Ellis:1998fv,Cordero:2009kv,Badger:2010mg,Frederix:2011qg,Oleari:2011ey} and was
the first in a set of off-shell five-particle amplitudes to be studied using the
unitarity method~\cite{Bern:1996ka,Bern:1997sc}. The present state of the art in phenomenological
studies allows full mass effects, shower matching, electro-weak corrections and the inclusion additional QCD jets~\cite{Luisoni:2015mpa,Kallweit:2014xda,Anger:2017glm}.

A numerical computation of the two-loop helicity amplitudes~\cite{Hartanto:2019uvl}
demonstrated the importance of an efficient analytic form with a well
understood basis of special functions. Major steps forward came via efficient numerical evaluation of the differential
equations~\cite{Abreu:2020jxa} and analytic evaluation in terms the
Goncharov Polylogarithms (GPLs)~\cite{Canko:2020ylt,Syrrakos:2020kba}. These results opened the
door for a fully analytic amplitude computation yet significant challenges
remain. The complexity of the external kinematics represents a challenge for
integral reduction techniques and the identification of a minimal basis of special
functions is required to find analytic simplifications after subtracting
universal infrared and ultraviolet divergences.

Efficient amplitude and integration-by-parts reduction (IBP)~\cite{Tkachov:1981wb,Chetyrkin:1981qh} using finite field arithmetic
\cite{Wang:1981:PAU:800206.806398,Wang:1982:PRR:1089292.1089293,Wang:1982:PRR:1089292.1089293,Trager:2006:1145768,RISC3778,vonManteuffel:2014ixa,Peraro:2016wsq,Peraro:2019svx,Smirnov:2019qkx,Klappert:2019emp,Klappert:2020nbg,Caola:2020dfu} has gained significant interest in recent years. Through multiple evaluations of a numerical algorithm~\cite{Badger:2017jhb,Abreu:2017hqn,Badger:2018gip,Abreu:2018jgq}, fully analytic forms for planar massless
five-particle amplitudes have been extracted using a rational parametrisation
of the kinematics~\cite{Hodges:2009hk}. Following a complete understanding of a pentagon function
basis~\cite{Gehrmann:2018yef,Chicherin:2020oor}, a large number of two-loop amplitudes are now available in compact
analytic form~\cite{Gehrmann:2015bfy,Badger:2018enw,Abreu:2018zmy,Abreu:2019odu,Abreu:2020cwb,Chawdhry:2020for,Abreu:2018aqd,Chicherin:2018yne,Chicherin:2019xeg,Abreu:2019rpt,Badger:2019djh,Agarwal:2021grm,DeLaurentis:2020qle}.
We have also seen the first phenomenological predictions at NNLO in QCD for the
production of three photons in hadron colliders after combination with real-virtual and
double real radiation~\cite{Chawdhry:2019bji,Kallweit:2020gcp}.

In this short letter we outline the extension of this method to processes with an additional mass scale.

\section{Leading colour \udtoWbb amplitudes \label{sec:amp}}

\begin{figure}[b]
  \begin{center}
    \includegraphics[width=0.4\textwidth]{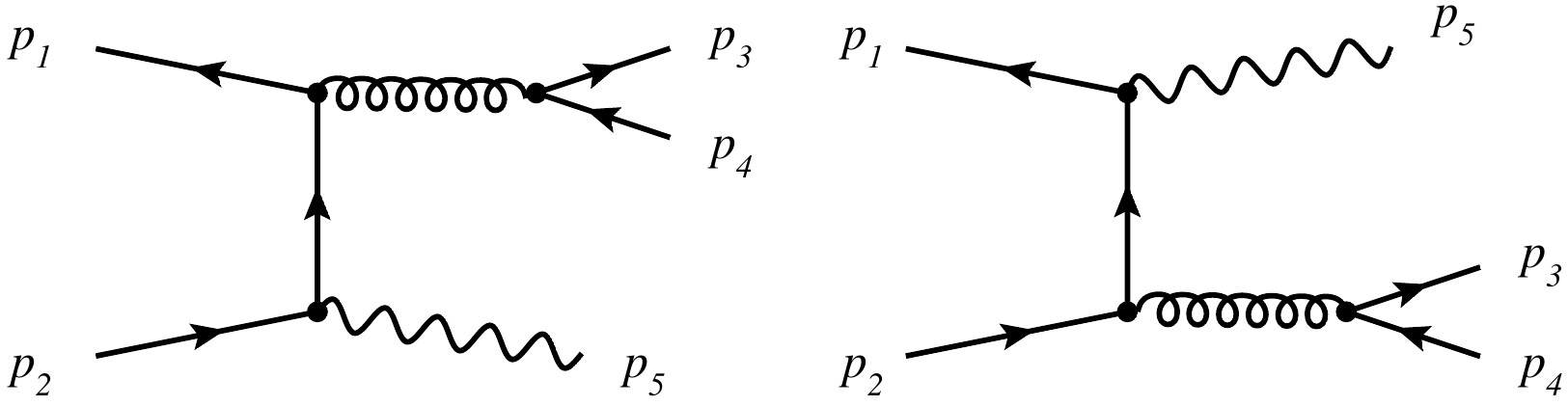}
  \end{center}
  \caption{Leading order Feynman diagrams contributing to \udtoWbb.}
  \label{fig:lodiags}
\end{figure}

\begin{figure}[b]
  \begin{center}
    \includegraphics[width=0.5\textwidth]{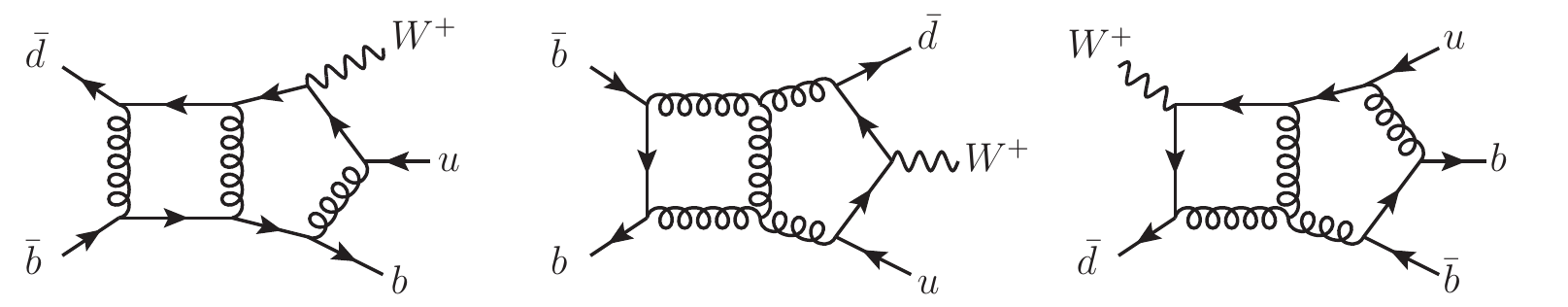}
  \end{center}
  \caption{Sample Feynman diagrams in the leading colour two-loop \udtoWbb amplitude.}
  \label{fig:samplediags}
\end{figure}

The leading order process consists of two simple Feynman diagrams as shown in Fig.~\ref{fig:lodiags}. We label our process as follows,
\begin{equation}
\bar{d}(p_1) + u(p_2) \rightarrow b(p_3) + \bar{b}(p_4) + W^+(p_5),
\end{equation}
where $p_1^2=p_2^2=p_3^2=p_4^2=0$ and $p_5^2=m_W^2$. The colour decomposition at leading colour is
\begin{align}
  & \mathcal{A}^{(L)}(1_{\bar{d}},2_{u},3_{b},4_{\bar b} ,5_{W}) =  \nn 
& \qquad n^L g_s^2 g_W  \delta_{i_1}^{\;\;\bar i_4} \delta_{i_3}^{\;\;\bar i_2}   A^{(L)}(1_{\bar d},2_{u},3_{b},4_{\bar b} ,5_{W}),
\label{eq:colourdecompositionW}
\end{align}
where $n= m_\eps N_c \alpha_s/(4\pi),\ \alpha_s = g_s^2/(4\pi)$ and $m_\eps=i (4\pi)^{\eps} e^{-\eps\gamma_E}$. $g_s$ and $g_W$ are the strong and weak coupling constants respectively.

We interfere the $L$-loop partial amplitudes $A^{(L)}$ in Eq.~\eqref{eq:colourdecompositionW} with the tree-level partial amplitude $A^{(0)}$
to obtain the unrenormalised $L$-loop unpolarised squared partial amplitude,
\begin{align}
 M^{(L)} =  \sum_{\mathrm{spin}} A^{(0)*}   A^{(L)}. 
\label{eq:squaredamp}
\end{align}
After the interference with the tree-level amplitude the analytic expression can be written in terms of scalar invariants,
\begin{align}
&s_{12} = (p_1 + p_2)^2\,, \quad s_{23} = (p_2 - p_3)^2\,, \quad s_{34} = (p_3 + p_4)^2\,, \nn
&s_{45} = (p_4 + p_5)^2\,, \quad s_{15} = (p_1 - p_5)^2\,, \quad s_{5} = p_5^2\,,
\label{eq:invariants}
\end{align}
and a parity-odd quantity, $\mathrm{tr}_5 = 4 i \eps_{\mu \nu \rho \sigma} p_1^{\mu} p_2^{\nu} p_3^{\rho} p_4^{\sigma}$.
Our results are the so-called finite remainders $F^{(L)}$, obtained after subtraction of infrared and ultraviolet divergences, $F^{(L)} = M^{(L)} - P^{(L)}$,
where $P^{(L)}$ takes the well known form~\cite{Catani:1998bh,Becher:2009qa,Becher:2009cu,Gardi:2009qi}. 
The explicit form for our process using the same conventions can be found in Ref.~\cite{Hartanto:2019uvl}.

\section{Amplitude reduction\label{sec:reduction}}

Feynman diagrams for the $u\bar{d}\rightarrow W^+ b\bar{b}$ scattering are generated using \textsc{Qgraf}~\cite{Nogueira:1991ex}.
In the leading colour approximation, there are 2, 16 and 210 diagrams contributing to the tree level, 1-loop and 2-loop amplitudes, respectively.
Example 2-loop diagrams are shown in Figure~\ref{fig:samplediags}.
Upon interference of the $L$-loop partial amplitude $A^{(L)}$ with the tree level partial amplitude $A^{(0)}$ according to Eq.~\eqref{eq:squaredamp}, 
the squared partial amplitude can be written as
\begin{equation}
M^{(L)}(\lrbrace{p})  = \int \prod_{i=1}^{L} \frac{d^d k_i}{i \pi^{d/2}e^{-\eps \gamma_E}}
\sum_{T}  \frac{N_T(d,\lrbrace{k},\lrbrace{p})}{\prod_{\alpha \in T}
D_\alpha(\lrbrace{k},\lrbrace{p})},
\label{eq:squaredampform}
\end{equation}
where $p$ are the external momenta which live in four dimensions, and $k_i$ are
the loop momenta. We work in the conventional
dimensional regularisation (CDR) scheme, where we have $d=4-2\eps$ dimensions.

The $W$-boson polarisation sum is performed in the unitary gauge,
\begin{equation}i
\sum_{\lambda} \wpol^{\mu *}(p_5,\lambda) \wpol^\nu(p_5,\lambda) = - g^{\mu\nu} + \frac{p_5^\mu p_5^\nu}{m_W^2}.
\end{equation}
The terms containing traces with a single $\gamma_5$ are treated using Larin's prescription~\cite{Larin:1993tq}, while
those with two $\gamma_5$'s are treated using the anti-commuting $\gamma_5$ prescription.
Larin's prescription has been employed in a wide variety of multi-loop computations and
a detailed discussion can be found, for example, in Ref.~\cite{Moch:2015usa}.
We have checked that using Larin's scheme throughout gives the same results for $F^{(L)}$.
We can then split the squared partial amplitude into parity-even and parity-odd parts,
\begin{equation}
M^{(L)}(\lrbrace{p}) = M_{\mathrm{even}}^{(L)}(\lrbrace{p}) + \mathrm{tr}_5 M_{\mathrm{odd}}^{(L)}(\lrbrace{p}).
\end{equation}
$M_{\mathrm{even}}^{(L)}$ receives contribution from the terms with no or two $\gamma_5$'s while $\mathrm{tr}_5 M_{\mathrm{odd}}^{(L)}$ is made up of terms with a single $\gamma_5$.
We note that the parity-odd part vanishes at tree level, $M^{(0)}_\mathrm{odd}=0$.

\begin{figure}[t]
  \begin{center}
    \includegraphics[width=0.5\textwidth]{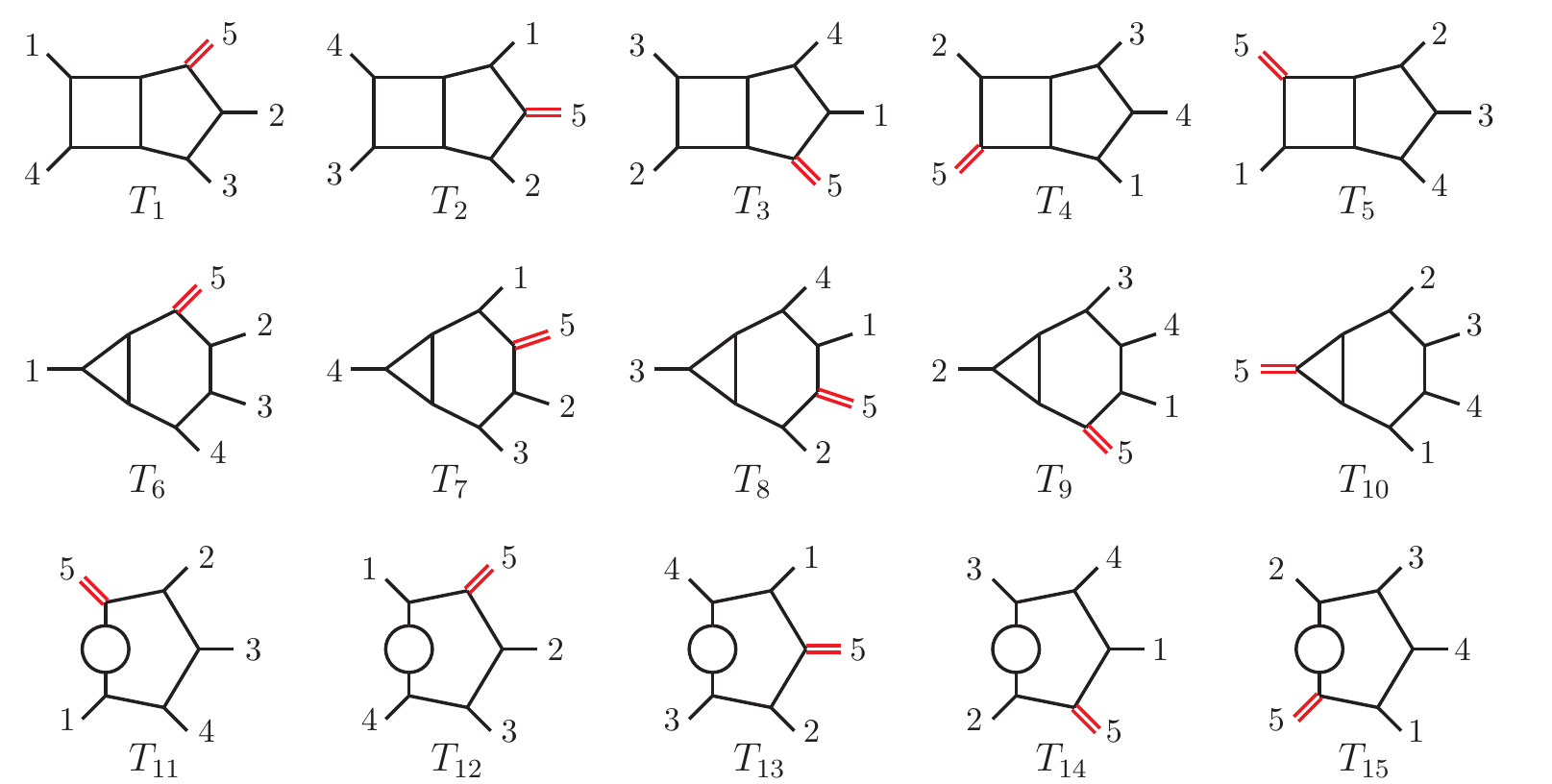}
  \end{center}
  \caption{Topologies with maximum number of propagators.}
  \label{fig:maxtopos}
\end{figure}

To perform the reduction of the 2-loop amplitude onto a basis of master integrals we first map each topology $T$ to a set of 15 maximal cut or \textit{master topologies} as shown in Fig.~\ref{fig:maxtopos}. The master topologies are then defined with a spanning set of 11 propagators and, after tracking shifts in the loop momentum, the change of variables for each topology $T$ can be computed. 
The resulting squared partial amplitude is now written as a linear combination of scalar integrals $\cI$
\begin{equation}
M_k^{(2)}(\lrbrace{p})  = \sum_{i} c_{k,i}(\eps,\lrbrace{p}) \; \cI_{k,i}(\eps,\lrbrace{p}),
\label{eq:squaredampIhat}
\end{equation}
where $k\in\lbrace\mathrm{even},\mathrm{odd}\rbrace$.
Analytic forms of the unreduced squared matrix elements above are derived using a collection of 
\textsc{Form}~\cite{Kuipers:2012rf,Ruijl:2017dtg} and \textsc{Mathematica} routines.
The integrals appearing in Eq.~\eqref{eq:squaredampIhat} are not all
independent. Relations between integrals $\cI$ can be found using IBP identities
and the squared amplitude can be written in terms of an independent set of
master integrals as follows
\begin{equation}
M_k^{(2)}(\lrbrace{p})  = \sum_{i} d_{k,i}(\eps,\lrbrace{p}) \; \mathrm{MI}_i(\eps,\lrbrace{p}).
\label{eq:squaredampMI}
\end{equation}
The reduction to master integral basis is then performed within the \textsc{FiniteFlow}
framework~\cite{Peraro:2019svx}, separately for $M_{\mathrm{even}}^{(2)}$ and $M_{\mathrm{odd}}^{(2)}$.
We use \textsc{LiteRed}~\cite{Lee:2012cn} 
to generate the IBP relations in \textsc{Mathematica}, together with the Laporta algorithm~\cite{Laporta:2001dd}
to solve them numerically over finite fields. 
We note that only master topologies $T_1 - T_{10}$ are included in the IBP system since 
the integrals belonging to master topologies $T_{11}-T_{15}$ can be mapped onto master topologies $T_{6}-T_{10}$.
The procedure for performing the reduction
onto master integrals using IBP relations is of course extremely well known,
the challenge in this example is one of enormous algebraic complexity. By
encoding the problem within a numeric sampling modular arithmetic we are able
to efficiently solve the Laporta system with tensor integral ranks of up to
five, avoiding all large intermediate expressions. For planar topologies such
as the ones appearing here the application of syzygy relations~\cite{Gluza:2010ws,Ita:2015tya,Larsen:2015ped} to optimise the
IBP reduction would likely lead to a substantial speed-up in computation time,
although in our case it was not found to be necessary. We did not perform an
analytic reconstruction after completing the set up of the reduction in
\textsc{FiniteFlow} graphs. Instead we continued to map the amplitude onto a
basis of special functions. 

\section{A basis of special functions for the finite remainder}

There are 202 master integrals contributing to the amplitude, 196 of them are covered by the 3 independent pentabox master integral topologies, while 6 are of one-loop squared type that involve one-loop massive on-shell bubble integral. We choose the canonical bases of master integrals constructed in Ref.~\cite{Abreu:2020jxa}. They satisfy differential equations (DEs)~\cite{Kotikov:1990kg,Bern:1993kr,Remiddi:1997ny,Gehrmann:1999as} in the canonical form~\cite{Henn:2013pwa},
\begin{align} \label{eq:canonicalDEs}
d \overrightarrow{\text{MI}} = \eps \sum_{i=1}^{58}a_i d\log w_i \, \overrightarrow{\text{MI}}  \, ,
\end{align}
where $\overrightarrow{\text{MI}}$ is the set of canonical master integrals for any of the involved topologies, the $a_i$ are constant rational matrices, while $\{w_i\}_{i=1}^{58}$ is a set of algebraic functions of the external kinematics called letters (see Ref.~\cite{Abreu:2020jxa} for their definition). The alphabet, i.e. the set of all letters, is the same for all planar one-mass five-particle integrals up to two loops, whereas the constant matrices $a_i$ depend on the topology. In Ref.~\cite{Abreu:2020jxa}, the authors also discuss a strategy to evaluate the master integrals numerically, based on the solution of the DEs~\eqref{eq:canonicalDEs} in terms of generalised power series~\cite{Francesco:2019yqt}. More recently, analytic expressions of the canonical master integrals in terms of GPLs~\cite{Goncharov:1998kja,Remiddi:1999ew,Goncharov:2001iea} have become available~\cite{Canko:2020ylt,Syrrakos:2020kba}. Both approaches allow for the numerical evaluation of the master integrals in any kinematic region and with arbitrary precision. Both approaches, however, also share certain drawbacks. Whether we reconstruct the prefactors of the $\eps$-components of the master integrals in Eq.~\eqref{eq:squaredampMI} or we map the latter onto monomials of GPLs, we cannot subtract the infrared and ultraviolet poles analytically and reconstruct directly the finite remainder. 

We overcome these issues by constructing a basis out of the $\eps$-components of the canonical master integrals up to order $\eps^4$. The crucial tool we employ in this construction are Chen's iterated integrals~\cite{Chen:1977oja}. We can define them iteratively through
\begin{equation} \label{eq:Chen}
\begin{aligned}
& d [w_{i_1}, \ldots, w_{i_n}]_{s_0} (s) = d\log w_{i_n} [w_{i_1}, \ldots,  w_{i_{n-1}}]_{s_0} (s) \, , \\
& [w_{i_1}, \ldots, w_{i_n}]_{s_0} (s_0) = 0 \, ,
\end{aligned}
\end{equation}
where $s$ denotes cumulatively the kinematic invariants, $s_0$ is an arbitrary boundary point, and the iteration starts from $[]_{s_0}(s) = 1$. The depth $n$ of the iterated integral is called transcendental weight. We refer to the notes~\cite{Brown:2013qva} for a thorough discussion. All GPLs can be rewritten in terms of iterated integrals. The latter however offer two useful advantages. The first is that --~conjecturally~-- they implement automatically all the functional relations. Once a GPL expression is rewritten in terms of iterated integrals in a given alphabet $\{w_i\}$, finding the functional relations becomes a linear algebra problem, as `words' $[w_{i_1},\ldots,w_{i_n}]$ with different letters are linearly independent. The second is that it is completely straightforward to write out the solution of the canonical DEs~\eqref{eq:canonicalDEs} in terms of iterated integrals. 
Eq.~\eqref{eq:canonicalDEs} in fact implies the following differential relation between consecutive components of the $\eps$ expansion of the master integrals,
\begin{align} \label{eq:canonicalDEsEps}
d \overrightarrow{\text{MI}}^{(k)} = \sum_{i=1}^{58} a_i d\log w_i \, \overrightarrow{\text{MI}}^{(k-1)}  \, , \quad \forall k\ge 1 \, ,
\end{align}
where $\overrightarrow{\text{MI}}^{(k)}$ is the $\mathcal{O}(\eps^k)$ term of the master integrals. Comparing Eq.~\eqref{eq:canonicalDEsEps} to Eq.~\eqref{eq:Chen}, we see that the iterated integral expressions of $\overrightarrow{\text{MI}}^{(k)}$ are obtained by adding a letter to the right of those of the previous order, multiplying them by the constant matrices $a_i$, and adding the boundary values. The master integrals are normalised to start from $\mathcal{O}(\eps^0)$ and so the $\mathcal{O}(\eps^k)$ components have transcendental weight $k$.

We used the GPL expressions of Refs.~\cite{Canko:2020ylt,Syrrakos:2020kba} to compute the values of the master integrals in an arbitrary point $s_0$ with $1100$-digit precision. Using the PSLQ algorithm~\cite{PSLQ}, we determined the integer relations among the boundary values, and rewrote them in terms of a basis of transcendental constants. Next, we used the differential equations provided by Ref.~\cite{Abreu:2020jxa} to express the relevant master integrals in terms of iterated integrals. This allowed us to determine a minimal set of linearly independent integral components, order by order in $\eps$ up to $\eps^4$. We denote these functions by $\{f^{(w)}_i\}$, where $w=1,\ldots,4$ labels the weight. Since each $f^{(w)}_i$ corresponds to an $\eps$ component of the master integrals, we can evaluate them numerically using the methods of Refs.~\cite{Abreu:2020jxa,Canko:2020ylt,Syrrakos:2020kba}, with the additional advantage that they are linearly independent. 

In order to subtract the poles analytically, we need to be able to write in the same basis also the 
subtraction term. From the transcendental point of view, the latter is given by the product of certain logarithms and transcendental constants coming from the anomalous dimensions --~$\pi^2$ and $\zeta_3$~-- times the one-loop amplitude. In order to accommodate this in the basis, we add the transcendental constants as elements, and work out the relations between the functions at each weight and products of lower-weight ones using the shuffle algebra of the iterated integrals. As a result, the functions in the basis $\{f^{(w)}_i\}$ are indecomposable, i.e. they cannot be rewritten in terms of lower-weight elements of the basis.

Armed with this function basis, we can proceed with the reconstruction of the two-loop finite remainders. We map the master integrals appearing in Eq.~\eqref{eq:squaredampMI} onto a monomial basis of the functions $\{f^{(w)}_i\}$, which we denote by $\{m(f)\}$, and perform a Laurent expansion in $\eps$ up to $\cO(\eps^0)$. We do the same for the subtraction term $P^{(2)}$. The resulting finite remainder,
\begin{equation}
F_k^{(2)}(\lrbrace{p})  = \sum_{i} e_{k,i}(\lrbrace{p}) \;  m_{k,i}(f) + \cO(\eps) \, ,
\label{eq:F2}
\end{equation}
is indeed free of $\eps$ poles. We set $s_{12}=1$ to simplify the reconstruction. The dependence can be recovered a posteriori through dimensional analysis.
The coefficients $e_{k,i}(\lrbrace{p})$ in $F_k^{(2)}$ are not all independent. We find the linear relations between them and a set of additional coefficients we supply as ansatz. We used tree-level expressions, coefficients from the one-loop amplitude and from the unreduced scalar integrals.
Through these linear relations we rewrite the complicated coefficients in $F^{(2)}$ in terms of known coefficients from the ansatz and simpler ones, which finally have to be reconstructed.
Moreover, we simplify the reconstruction of the remaining coefficients by partial fractioning them with respect to $s_{23}$. First we determine the denominator factors by computing a univariate slice and matching it against an  ansatz made of letters $w_i$. Using the information about the denominator and the polynomial degree in the numerator, we construct an ansatz for the partial-fractioned expressions of the coefficients. Then we fit the ansatz with a numerical sampling. See Refs.~\cite{Abreu:2019odu,Boehm:2020ijp,Heller:2021qkz} for recent work on multivariate partial fractioning. To emphasize the effectiveness of our strategy, we note that the coefficients of the parity-even (-odd) two-loop amplitude written in terms of GPL monomials have maximal degree $62$ ($63$). The maximal degree drops to $54$ ($54$) when we use the basis of special functions $\{f^{(w)}_i\}$ in the finite remainder, and then to $31$ ($32$) in the remaining $4$ variables after partial fractioning. The reconstruction finally required $38663$ ($45263$) sample points over $2$ prime fields, gaining a factor of $7$ in the reconstruction time with respect to the GPL-based approach~\footnote{Since estimates of evaluation time rely on system specific parameters, we have taken a conservative value for the speed improvement.}.
The reconstructed analytic expressions are further simplified using the \textsc{MultivariateApart} package~\cite{Heller:2021qkz}.

The iterated integrals expression of the $f^{(w)}_i$ functions allow us to study the analytic structure of the finite remainder in a very convenient way. Interestingly, we observe that certain letters do not appear. As it was already noted in Ref.~\cite{Abreu:2020jxa}, the last 9 letters do not show up in any two-loop amplitude up to order $\eps^0$. Out of the relevant 49 letters, 6 ($w_i$ with $i\in\{16,17,27,28,29,30\}$) appear in the master integrals but cancel out in the two-loop amplitude truncated at $\cO(\eps^0)$. Finally, the letter $w_{49}=\mathrm{tr}_5$ is present in the two-loop amplitude, but cancels out in the finite remainder. This letter has already been observed to exhibit the same behaviour in all the known massless two-loop five-particle amplitudes~\cite{Badger:2018enw,Abreu:2018zmy,Abreu:2018aqd,Chicherin:2018yne,Chicherin:2019xeg,Abreu:2019rpt,Abreu:2019odu,Badger:2019djh,Caron-Huot:2020vlo,Abreu:2020cwb,Chawdhry:2020for}, which has spawned interest in the context of cluster algebras~\cite{Chicherin:2020umh}.

As regards the numerical evaluation, we propose a strategy based on the generalised power series solution of the DEs~\cite{Francesco:2019yqt} applied not to the master integrals, but directly to the basis of special functions. If we rescale each special function in the basis $f^{(w)}_i$ by a power of $\eps$ corresponding to its weight, the ensuing list of functions $\vec{v} = \{\eps^w f^{(w)}_i, 1 \}$ satisfies a system of DEs in the canonical form~\eqref{eq:canonicalDEs}. This follows from the differential property of the iterated integrals~\eqref{eq:Chen}. Differently from the DEs for the master integrals, the DEs for the special functions contain only the minimal amount of information necessary to evaluate the finite remainder. For instance, instead of evaluating all the weight-4 functions that may appear in any one-mass two-loop five-particle amplitude, we can restrict ourselves to evaluating only the 19 linear combinations that actually appear in $F_k^{(2)}$. We therefore define a new basis of special functions, $\{g^{(w)}_i\}$, which at weight four includes the aforementioned 19 combinations of $f^{(4)}_i$'s, at weight three contains only the $f^{(3)}_i$'s appearing in $F_k^{(2)}$ and in the derivatives of the $g^{(4)}_i$'s, and so on down to weight zero. The resulting DEs are much simpler that those for the master integrals. For instance, they are by-construction free of the letters which do not appear in the finite-remainder, and their dimension is smaller than the number of master integrals for all the relevant families. Finally, we evaluate the $g^{(w)}_i$'s by solving the corresponding DEs using the \textsc{Mathematica} package \textsc{DiffExp}~\cite{Hidding:2020ytt}. We compute the boundary values in an arbitrary point in the physical scattering region through the correspondence between the $g^{(w)}_i$'s and the master integral components. 

The complete analytic expression of the two-loop finite remainder in terms of rational coefficients and special functions is included in the ancillary files, together with the differential equation and the boundary values necessary to evaluate the latter numerically. 
We performed Ward identity checks at the level of master integrals for $M_{\mathrm{even}}^{(2)}$ and at the level of the finite remainder for $M_{\mathrm{odd}}^{(2)}$: 
we modified the numerator functions by replacing the loop and tree-level amplitude polarisation vectors with $p_5$ and $p_1$ respectively, and observed that $M_{\mathrm{even}}^{(2)}$ and $F_{\mathrm{odd}}^{(2)}$ vanish.
We also compared numerically the finite remainders derived in this work against results from an independent helicity amplitude computation 
in the t'Hooft-Veltman scheme using the framework of Ref.~\cite{Hartanto:2019uvl}.
For the convenience of future cross-checks, we provide the numerical values of $M_k^{(2)}$ and $F_k^{(2)}$ at one phase space point in Table~\ref{tab:benchmark2L}. 

\renewcommand{\arraystretch}{1.5}
\begin{table}[t!]
\centering
\begin{tabularx}{0.45\textwidth}{|C{0.4}|C{1.3}|C{1.3}|}
\hline
  & Re$\lbrace M_{\mathrm{even}}^{(2)}/M^{(0)}_{\eps=0}\rbrace$ & Re$\lbrace\mathrm{tr}_5 M_{\mathrm{odd}}^{(2)}/M^{(0)}_{\eps=0}\rbrace$ \\
\hline
$\eps^{-4}$ &   2  & 0 \\
$\eps^{-3}$ &  -2.19718713546 & 0 \\
$\eps^{-2}$ & -12.7892676147  & -0.211614995129 \\
$\eps^{-1}$ & -7.77698255746  &   11.5990058250 \\
$\eps^{0}$  &  116.073111075  &   28.7730449523 \\
\hline 
  & Re$\lbrace F_{\mathrm{even}}^{(2)}/M^{(0)}_{\eps=0}\rbrace$ & Re$\lbrace\mathrm{tr}_5 F_{\mathrm{odd}}^{(2)}/M^{(0)}_{\eps=0}\rbrace$ \\
\hline  
$\eps^{0}$  &  144.141227186  & -7.34974777490 \\
\hline
\end{tabularx}
\caption{\label{tab:benchmark2L} Numerical results for the leading colour two-loop squared partial amplitude, $M_k^{(2)}$, and 
finite remainder, $F_k^{(2)}$, normalised to the tree level squared partial amplitude in 4 dimensions, $M^{(0)}_{\eps=0}$,
at the physical point $\{s_{12}=5, s_{23}=-1/3, s_{34} = 11/13, s_{45} = 17/19, s_{15} = -23/29, s_5 = 1/7\}$.
}
\end{table}

\section{Discussion and Outlook\label{sec:results}}

\begin{figure}[t]
  \begin{center}
    \includegraphics[width=0.4\textwidth]{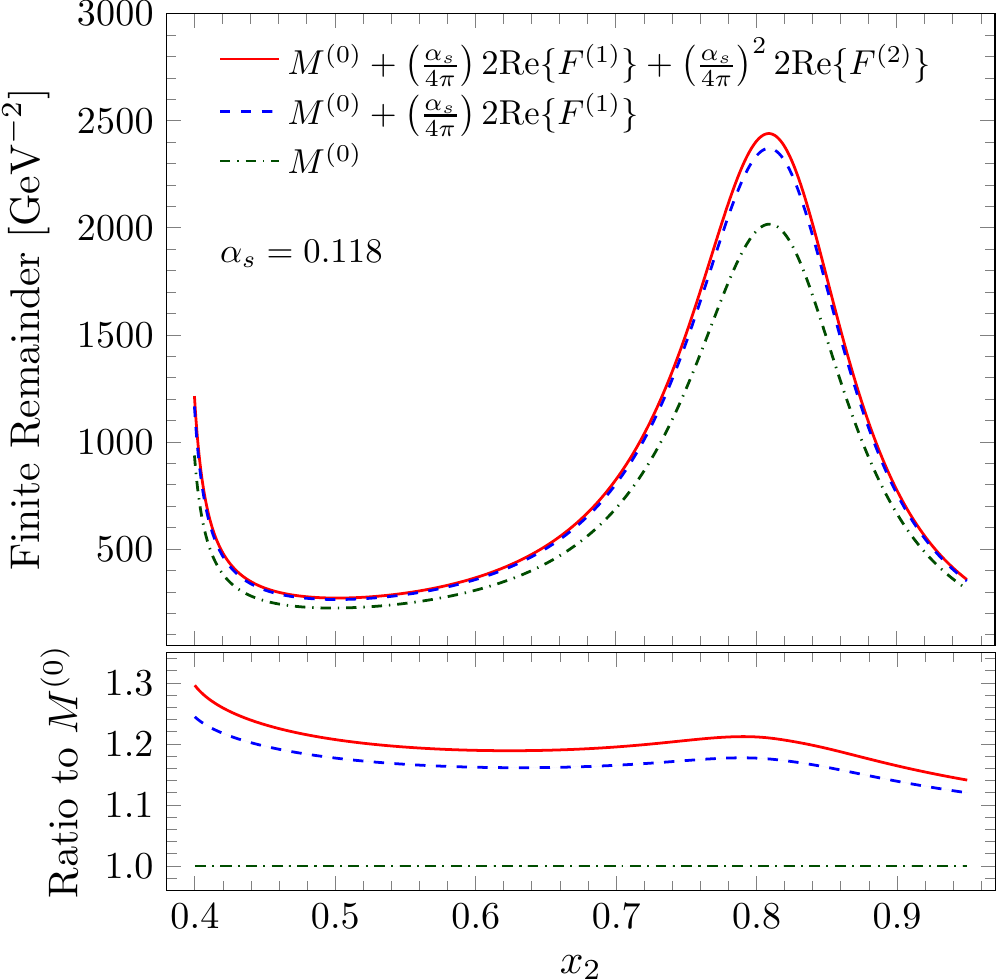}
  \end{center}
  \caption{The finite remainder $F^{(L)} = F_\mathrm{even}^{(L)} + \mathrm{tr}_5 F_\mathrm{odd}^{(L)}$ for \udtoWbb 
at one and two loops as a function of the variable $x_2$ defined in Eq.~\eqref{eq:parametrisation}.}
  \label{fig:ampplot}
\end{figure}

The results we have obtained represent a major step forward and open the door
to phenomenological applications. The identification of a basis of special
functions has resulted in a substantial speed up over previous studies as well
as uncovering explicit cancellations and reduction in complexity. To
demonstrate the suitability for phenomenological applications we present the
evaluation on a univariate slice of the physical phase space. For this we use a
parametrisation in terms of energy fractions and angles of the final state,
\begin{equation} \label{eq:parametrisation}
\begin{aligned}
  p_3 &= \tfrac{x_1\sqrt{s}}{2} \left( 1, 1, 0, 0\right) \, , \\
  p_4 &= \tfrac{x_2\sqrt{s}}{2} \left( 1, \cos\theta, -\sin\phi\sin\theta,  -\cos\phi\sin\theta\right) \, , \\
  p_5 &= \sqrt{s} \left( 1, 0, 0, 0 \right)-p_3-p_4 \, ,
\end{aligned}
\end{equation}
where $p_1$ and $p_2$ are taken back-to-back along the $z$-axis with a total
centre-of-mass energy of $s$. We have chosen $p_3$ to be produced at an
elevation of $\tfrac{\pi}{2}$ from the $z$-axis and the on-shell phase space
conditions impose $\cos\theta = 1 + \tfrac{2}{x_1 x_2}\left( 1-x_1-x_2 - \tfrac{m_W^2}{s}\right)$.
In Fig.~\ref{fig:ampplot} we plot values of the one- and two-loop finite
remainders against $x_2$ for a configuration with $\phi=0.1, m_W=0.1, s=1$ and
$x_1=0.6$.  The special functions were evaluated with
\textsc{DiffExp}~\cite{Hidding:2020ytt} using rationalized values of the
invariants. An average evaluation time of $260$s over $1000$ points was
observed and the function is smooth and stable over the whole region. This
demonstrates that even with a basic setup in \textsc{Mathematica} a reasonable
evaluation time can be achieved and that realistic phenomenology can now be
performed.

The results obtained here pave the way for a
broader class of $2\to3$ scattering problems. The solution of the IBP system	
and the basis of special functions do not depend on the on-shell approximation
of the $W$-boson and apply equally to the planar sectors of $pp\to W/Z+2j$
(including decays) and $pp\to H+2j$. Going beyond leading colour for $pp\to
W/Z+2j$ or any complete $pp\to H+2j$ amplitudes at two-loops still requires
missing information on the non-planar master integrals, nevertheless we believe
they can be easily incorporated into the strategy we introduce here.

\appendix

\begin{acknowledgments}
We thank Herschel Chawdhry, Thomas Gehrmann, Johannes Henn, Alexander Mitov, Tiziano Peraro and Rene Poncelet for numerous insightful discussions. 
We also thank Nikolaos Syrrakos for kindly providing the results of Ref.~\cite{Syrrakos:2020kba} prior to its publication.
This project has received funding from the European Union’s Horizon 2020 research and innovation programmes
\textit{New level of theoretical precision for LHC Run 2 and beyond} (grant agreement No 683211),
\textit{High precision multi-jet dynamics at the LHC} (grant agreement No 772009), and
\textit{Novel structures in scattering amplitudes} (grant agreement No 725110).
HBH has been partially supported by STFC consolidated HEP theory grant ST/T000694/1.
SZ gratefully acknowledges the computing resources provided by the Max Planck Institute for Physics.
\end{acknowledgments}

\bibliography{w4q_planar}

\end{document}